\documentclass[fleqn,usenatbib]{mnras}

\usepackage{newtxtext,newtxmath}

\usepackage[T1]{fontenc}

\DeclareRobustCommand{\VAN}[3]{#2}
\let\VANthebibliography\thebibliography
\def\thebibliography{\DeclareRobustCommand{\VAN}[3]{##3}\VANthebibliography}

\usepackage{graphicx}	
\usepackage{multirow}
\usepackage{float}

\title[Short title, max. 45 characters]{Detection of Einstein Telescope gravitational wave signals from binary black holes using deep learning}

\author[Wathela Alhassan et al.]{
Wathela Alhassan,$^{1}$\thanks{E-mail: wathelahamed@gmail.com}
T. Bulik,$^{1,2}$
M. Suchenek$^{1}$
\\
$^{1}$Particle Astrophysics Science and Technology Centre, Nicolaus Copernicus Astronomical Center, Rektorska 4, 00-614 Warsaw, Poland \\
$^{2}$Astronomical Observatory, University of Warsaw, Aleje Ujazdowskie 4, 00-478 Warsaw, Poland
}

\date{Accepted XXX. Received YYY; in original form ZZZ}

\pubyear{2022}

\begin{document}
\label{firstpage}
\pagerange{\pageref{firstpage}--\pageref{lastpage}}
\maketitle

\begin{abstract}
The expected volume of data from the third-generation gravitational waves (GWs) Einstein Telescope (ET) detector would make traditional GWs search methods such as match filtering impractical. This is due to the large template bank required and the difficulties in waveforms modelling. In contrast, machine learning (ML) algorithms have shown a promising alternative for GWs data analysis, where ML can be used in developing semi-automatic and automatic tools for the detection and parameter estimation of GWs sources. Compared to second generation detectors, ET will have a wider accessible frequency band but also a lower noise. The ET will have a detection rate for Binary Black Holes (BBHs) and Binary Neutron Stars (BNSs) of order ${10}^{5} - {10}^{6} \text{ year}^{-1}$ and ${7 \times {10}^{4}} \text{ year}^{-1}$ respectively. We explored the efficiency of using convolutional neural networks (CNNs) for the detection of BBHs’ mergers in synthetic noisy data that was generated according to ET’s parameters. Without performing data whitening or applying bandpass filtering, we trained four CNN networks with the state-of-the-art performance in computer vision, namely VGG, ResNet and DenseNet. ResNet has significantly better performance, and was able to detect BBHs sources with SNR of 8 or higher with 98.5\% accuracy, and with 92.5\%, 85\%, 60\% and 62\% accuracy for sources with SNR range of 7-8, 6-7, 5-6 and 4-5 respectively. ResNet, in qualitative evaluation, was able to detect a BBH's merger at 60 Gpc with 4.3 SNR. It was also shown that CNN can be used efficiently for near-real time detection of BBHs.
\end{abstract}




\section{Introduction}

The recent detection of Gravitational Waves (GWs) has opened a new era of multimessenger astrophysics, where significant amount of data is being delivered from the second generation detectors of GWs such as GEO600 \citep{Willke2007}, the advanced Laser Interferometer Gravitational-Wave Observatory (LIGO) \citep{Abbott2009}, and the Virgo interferometer \citep{Acernese2007}.




The detection of GW signals is traditionally done through the analysis of a hidden signal in the raw data from the detector, so-called matched filtering. Matched filtering process compares modelled waveform templates with the raw data. The main disadvantage of match filtering is that it depends entirely on the availability of waveform templates. This process is computationally challenging, and depends also on the accuracy of the modelled waveforms. Once a signal is detected, the subsequent step is to estimate its parameters, for which as many as ${10}^{7}$ waveform templates are required for an accurate estimation \citep{Schmidt_2020}.


With the expected volume of data from the third generation GWs detectors such as the Einstein Telescope (ET) \citep{Punturo_2010, abernathy2011einstein, 2020JCAP...03..050M}, Cosmic Explorer (CE) \citep{2021arXiv210909882E} and The Laser Interferometer Space Antenna (LISA) \citep{Shaddock_2008}, match filtering can be extremely computational expensive and the large template bank required would make it impractical. This motivates the necessity for the development of semi-automatic and automatic machine learning algorithms for detection, denoising and parameter estimation of GWs sources. Deep Learning (DL) is a promising alternative for GW data analysis, both for detection and parameter estimation. This goes back to its flexibility in categorizing and modelling patterns read from data, with very little human input. In contrast to match filtering search method, DL models can learn to distinguish between only noise and injected noise (noise plus signal) without the need to model every possible GW waveform. This is done through learning the underlying features of pure noise and injected noise, which gives DL an advantage over the traditional methods, where no matching is involved. DL algorithms can open up a new door for unmodelled waveforms discovery, specially with the expected number and type of sources that are expected to be detectable from the third generation detectors.


The ET will revolutionise the field of GWs in every sense, including and most importantly the better sensitivity of one order of magnitude that will provide, which will allow the detection at lower frequency. Compared to second generation detectors, ET will have a wider accessible frequency band that can be used to investigate a huge number of key issues related to astrophysics, fundamental physics and cosmology. The ET will have a detection rate for BBHs and BNSs of order ${10}^{5} - {10}^{6} \text{ year}^{-1}$ and ${7 \times {10}^{4} \text{ year}^{-1}}$ respectively \citep{2021arXiv211106990K}. A recent study showed that, the BBHs would have lower detection rate at ET, which expected to be in the range $3.1 \times {10}^{4}$ to $8.0 \times {10}^{4} \text{ year}^{-1}$ \citep{2022arXiv220702771I}. Our work here will focus on BBHs sources, and BNSs will be included in future work.

Over the last few years, different Machine Learning (ML) and DL techniques have been successfully applied  to astronomical applications, on both images and spectral data \citep{Lukic2016}. For example distinguishing between normal and peculiar optical galaxies by training Artificial Neural Networks (ANNs) \citep{Naim1996}. Estimation of the photometric redshift of galaxies using ANNs \citep{Collister2003}, improved methods of the estimation of the photometric redshift of galaxies with more accurate results \citep{Zhang2008, Bonfield2009} and most recent improvement by \cite{DIsanto2017}. \cite{Selim2016} and \cite{Selim2017} where a method based on non-negative matrix factorization algorithm were used to classify the galaxy morphology. Similar method has also been applied to the galaxy-star classification \citep{Kim2016}, and Radio sources classification using CNNs \cite{Alhassan2018}.

In GW, ML and DL have been wildly introduced into the field of GW data analysis in recent years. For second generation detectors ML and DL techniques have been successfully applied on both, time and frequency domain GWs data. For example, ML was used for glitches identification \citep{PhysRevD.101.102003}.\cite{PhysRevD.88.062003} used artificial neural networks, support vector machines, and random forests on LIGO data  for the identification and removal of glitches.
Another successful application used ML to discover the correlation between auxiliary monitoring channels and the main GWs output channels, for improving the the sensitivity of the current interferometric GWs detectors \citep{2022PhRvD.105j2005V}. The potential of time-domain noise subtraction for improving detectors sensitivity below 30 Hz was also discussed in the paper.
\cite{2021PhRvD.103j2003B} used CNNs for pre-merger  alert of three classes of BNSs of LIGO. This work was improved in \citep{2022arXiv220504750B} to be able to search for all types of BNSs pre-mergers in three different types of noise: Gaussian O3 noise, real O3 noise, and predicted O4 noise. \cite{2022MLS&T...3b5001M} used deep recurrent autoencoders (AEs) -with an unsupervised strategy- for GWs signals detection in laser interferometers. The results were better compared to other AEs architectures, but less accurate compared to the traditional supervised techniques.
\cite{2021CQGra..38o5010A} Used CNNs for BBHs merger detection from simulated data of a single detector (Advanced LIGO detector) and multi-detectors combined. Their models were trained on sources with luminosity distance ($D_{L}$) ranging from 100 to 2000 Mpc, and masses between 5 to 100 \(M_\odot\). Results from model trained on single detector data and multi-detector data achieved an accuracy of 72\% and 82\% respectively. Similar work was done by \cite{2022FrPhy..1754501J}, where CNNs was used to search the LIGO O1, O2 and O3 data for BBHs sources that have been detected by both, Hanford and Livingston detectors, using time sliding method. Their model was able to detect the whole 41 BBHs sources exist in the data.


Detection of Extreme mass ratio inspirals (EMRIs) can be possible from third generation space-borne GW detectors such as LISA. A proof-of-principle approach for EMRI detection based on CNNs was presented in \citep{PhysRevD.105.123027}, where their CNNs model was able to detect effectively EMRI systems with a signal-to-noise ratio (SNR) larger than 50. it has also been showed that, the CNNs' performance is strongly associated with SNR.

ET is expected to have a detection horizon for BBHs up to redshift z = 100, but BBHs with total mass 20 - 100 \(M_\odot\) will be visible up to $z \approx 20$ \citep{Singh:2021zah}. In this work, we focus on the detection of BBHs merger from the ET with luminosity distance up to 120 Gpc and component masses ranging from 15 to 56 \(M_\odot\). Different state-of-the-art CNNs models in image classification were used to find the most optimal performing model. Time-frequency domains of pure noise and injected noise have been used to train, validate and test our models.

The structure of this work goes as follow: in Section \ref{sec:data} we describe the method and tools were used for the generation of our synthetic data and the prepossessing pipeline employed for spectrogams preparation. Section \ref{sec:cnn} discusses The Convolutional Neural Networks, the DL networks used, and training and evaluation processes. Evaluating the best performing model on five hours of ET synthetic data is presented in section \ref{sec:eval}. Finally, a summary of this work and a future work plans is presented in Section \ref{sec:con}.

\section{ET dataset generation}
\label{sec:data}
Deep learning algorithms are data-hungry algorithms. This means they required a vast amount of data in order to generalize well. In our case -supervised classification- we have to provide thousands of samples to train and evaluate the algorithm. Our main task is to train different DL models, on frequency-time domain representations of only noise and injected noise, to be able to distinguish between the noise with a presence of GWs and without.

Time-frequency analysis, which is the prevalent method in distinguishing between GWs signal and noise, was used in the GWs discovery in finding the 'chirp' signal \citep{PhysRevLett.116.061102}. In this work, we use time-frequency domain (spectrograms) for the detection of GWs analysis, using The Short Time Fourier Transform (STFT). STFT is one of the most commonly used time-frequency transform for nonstationary signal analysis, known for its simplicity and efficiency, and it was also used in the current data analysis pipeline of LIGO \citep{PhysRevD.93.042004}. STFT was first introduced by Dennis Gabor in 1946, and it can mathematically be represented for a signal $x(t)$ as:
\begin{align*}
 X(\tau,\omega) = \int_{-\infty}^\infty  x(t)w(t-\tau) e^{-i\omega t} d(t) 
\end{align*}
Where $w$ is the window function, $\tau$ and $\omega$ are the time and frequency axis respectively. Choosing a window size is critical for the time-frequency resolution when using STFT \citep{6473982}. The main challenge in choosing a window size is the time-frequency localization trade-off (i.e. smaller window size will result in a better temporal resolution but lower frequency resolution). Through experiments, as a window, we chose $blackman$ window \citep{6768513} -which was designed to have the minimal possible spectral leakage that's is linked to STFT- with a length of 1024 and $50\%$ overlap resulting in a 62.5 ms temporal resolution. Figure \ref{fig:spec_samples} shows samples of the generated spectrograms.

ET is expected to observe individual sources, overlapped sources as well as stochastic background. We considered here only individual sources with no spinning. LALSimulation through pyCBC python package \citep{alex_nitz_2022_5825666}, using a single detector of the proposed ET telescope, was used for the generation of our GWs signals of BBHs and noise according to the ET's parameters. Combining data from different detectors can significantly improve the detection performance, compared to using a single detector as shown previously in \citep{2021CQGra..38o5010A}. We expect that, using all ET's three detectors would significantly improve the performance of our models, and this will be tried in future work.

Generation of ET’s noise using pyCBC incorporates two steps; the first step is the generation of power spectrum density (PSD) for ET, given the starting frequency ($f_{low}$) and frequency steps ($f_\Delta$). Second step is to generate a gaussian noise with a colour that matches the ET’s PSD, sampled at the ET’s frequency.

To generate GWs of a BBH sources, some parameters - of the source to be simulated - have to be provided, most importantly the masses of BBH and its $D_{L}$). These tow parameters - for all our sources - were taken from \cite{Singh:2021zah}, which contains a physical properties (masses and redshift $z$) of three populations of merging binaries (NSs and BHs) simulated specifically for the ET, using the ET-D design \citep{Hild:2010id}. To obtain $D_{L}$ from redshift $z$, we used Planck 2015 cosmology \citep{2016A&A...594A..13P} assuming flat LCDM, with Hubble constant, $H_0$= (67.8 +/- 0.9) $km/s/Mpc$, a matter density parameter $\Omega_m$ = 0.308 +/- 0.012, and a scalar spectral index with $n_s$ = 0.968 +/- 0.006.


pyCBC also requires specifying the approximant, which is the model to be used for the simulation. We used $IMRPhenomD$ approximant \citep{Husa2015} which describes time and frequency domain GWs from non-precessing BBHs. More parameters as in Table \ref{tab:param} was to be provided, such as Positions in the sky, time steps ($t_\Delta$), starting frequency ($f_{low}$), Inclination angle ($I_\theta$) , Polarization ($P_\delta$) and Coalesence ($C_\delta$) phases. It must be noted here that, lower starting frequency will result in longer lasting waveforms that can last up to days on the detector. Since we are interested in detecting the BBHs' mergers, starting frequency was set to 30 Hz to have shorter waveforms.

\begin{table}
\centering
\begin{tabular}{l|r}
PyCBC's parameter & Value \\\hline

Detector &	E1 detector\\
RA and Dec & Random (uniform distribution)\\
M1 and M2 & $\in$ 15-56 \(M_\odot\) \\ 
$D_{L}$ & 145-120,000 Mpc\\
$f_{low}$ & 30 Hz\\
$f_\Delta$	& 1.0/16\\
$t_\Delta$	& 1.0/16384\\
$I_\theta$ & cos ($\theta$) = uniform[0, $\pi$]\\
$P_\delta$	& uniform[0, 2$\pi$]\\
$C_\delta$ & uniform[0,  2$\pi$]

\end{tabular}
\caption{\label{tab:param}Parameters for generating GWs BBH sources.}
\end{table}

\subsection{Data processing}
The output of the pyCBC simulation is a one channel time-domain data (d(t), where d(t) is either a pure gaussian noise (n) or has a signal (h) been injected into it:

\begin{equation}
d(t) = 
    \begin{cases}
        h(t) + n(t), &\text{if data contains a BBH's signal}\\
        n(t), &\text{if data contains noise only}.
    \end{cases}
\end{equation}

Using the parameters in Table \ref{tab:param}, 25,000 BBH signals were generated with $D_{L}$ up to 120 Gpc and five SNR ranges: 4-5, 5-6, 6-7, 7-8 and >8. In ML, it's important to make sure the training data-set is well balanced, to make sure that classifiers won't be biased, in our case, towards a certain SNR range. To make sure that each SNR range has the same number of sources, we generated 180,000 BBH's signals and then divided them -based on their SNR values- into the five mentioned ranges. From each SNR range, 5000 signals were selected.

\begin{figure}
	\includegraphics[width=\columnwidth]{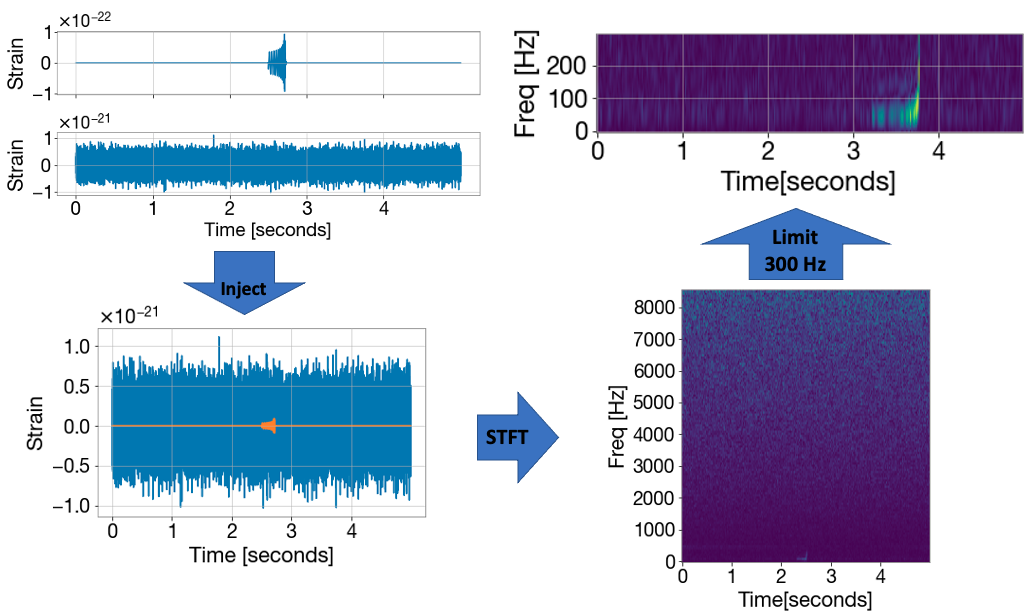}
    \caption{Data preparation and processing.}
    \label{fig:data_prep}
\end{figure}

\begin{figure}
\includegraphics[width=\columnwidth]{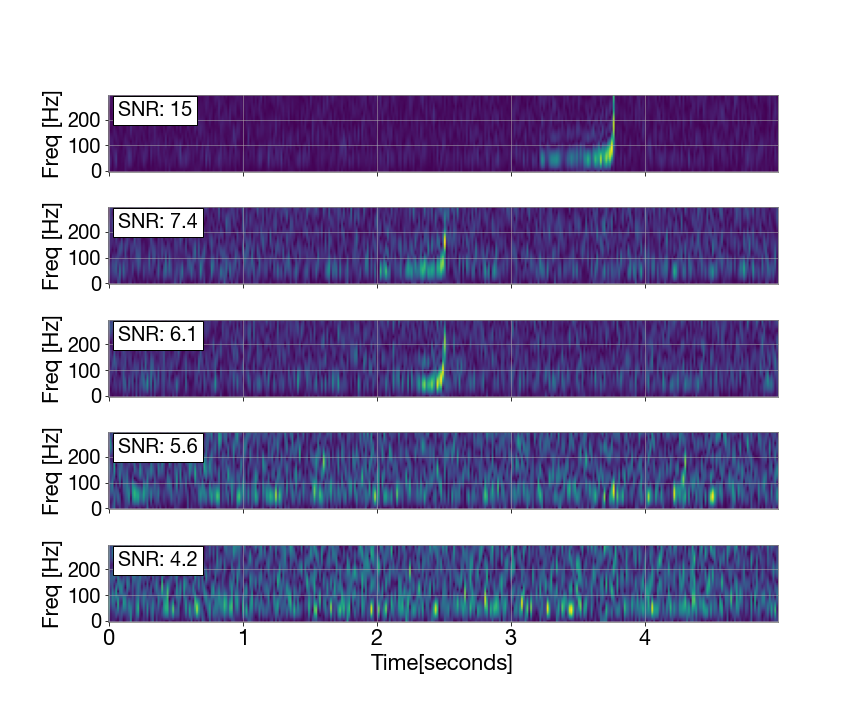}\\
\includegraphics[width=\columnwidth]{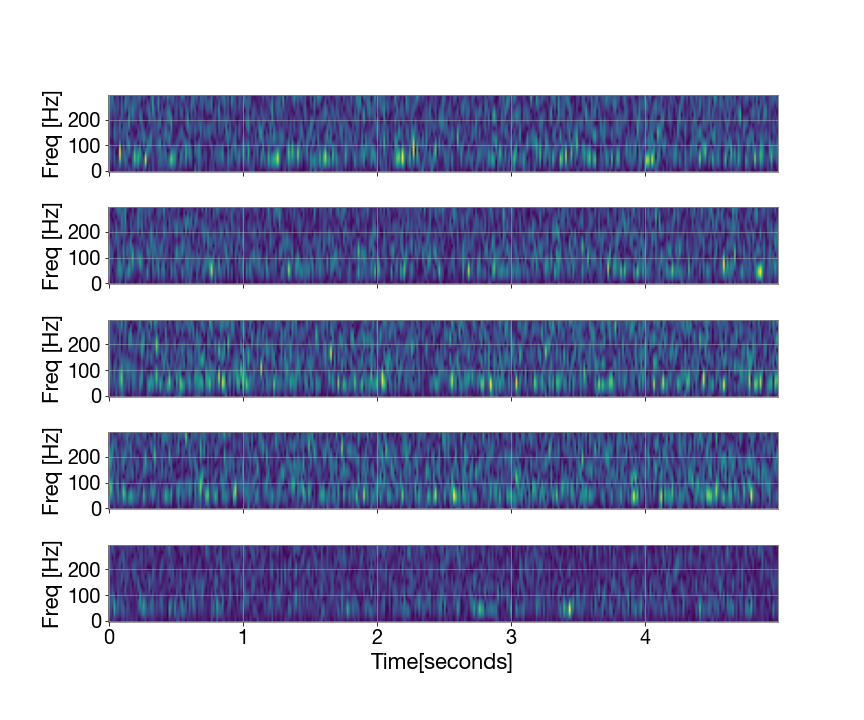}
\caption{Injected spectrogram samples: top, and only noise spectrogram samples: bottom.}
\label{fig:spec_samples}
\end{figure}

The data preparation and processing has four main steps as shown in Figure \ref{fig:data_prep}: (1) Signal and noise generation, generated signals has an observation time T ranges between 0.29 to 0.93 seconds, and each generated noise is a 5 seconds long. (2) Each signal was injected -in a random position- into the 5 seconds long noise. (3) STFT was used to produce the phase and magnitude spectrograms of each injection (noise plus signal) and a noise only time series. The result of this step is a spectrograms of only noise data (here after only noise spectrograms) and spectrograms of data with injected signals ( here after injected spectrograms). (4) Limit the frequency to 300 Hz, this is due to the fact that, our signals are always at the bottom of the spectrograms, lies within the 300 Hz limit. This step saves time and computational power needed for the training since we are removing a significant part of the spectrogram where it's only noise. Figure \ref{fig:spec_samples} displays samples of only noise and injected spectrograms, where we show samples of 5 injected signals with SNR belong to the five ranges. We did not whiten the data and no bandpass filtering  was applied. For the training of our CNN models, only the magnitude of the STFT was used and the complex part was ignored.
\section{Convolutional Neural Networks (CNN)}
\label{sec:cnn}
The Convolutional Neural Networks (ConvNet; \cite{Fukushima1980}) is a type of feed-forward neural network deep learning model. A feed-forward model means the output from one layer will be used as an input to the subsequent layer. CNNs are primarily designed to take advantage of the 2D structure of an input image (or other 2D input such as a speech signal) to solve complex tasks such as pattern recognition and object detection \citep{2015arXiv151108458O}. CNNs have a multi-layer structure, which contains different layers with different rules. These layers namely are, convolutional layers, pooling layers and fully-connected layers, for details please see  \citep{2014arXiv1404.7828S}. The main rule of a convolutional layer, is to extract features from the input (an image of a spectrogram in our case) through applying a user-defined number of filters, using a convolution operations \citep{Goodfellow-et-al-2016}. Each convolutional layer is usually followed by an activation function -to introduce none-linearity in the model- such as Hyperbolic Tangent $tanh$, $sigmoid$ \citep{10.5555/521706}, $softmax$ \citep{10.1007/978-3-642-76153-9_28}, or Rectified Linear Activation function ($ReLU$) \citep{2018arXiv180308375A}. A Pooling layer comes after the activation function and its main rule is to apply an scalar to vector transformation on a local regions of an image, to generate a representative value of the pixels in that region. Once the features have been pooled, it will be then flattened and fed to a fully connected layer. The rule of the fully-connected layer is same as Artificial Neural Networks (ANNs) which acts as a features classifier \citep{2017arXiv171203541A}. CNNs are essentially made up of neurons that receive inputs. When training an CNNs model, the model will learn the associated features of each class, through minimizing the difference (loss) between output prediction and training true labels. For details please see \citep{2015arXiv151108458O}. In short, the CNNs process describes a none-linear mapping between the input images and their associated class scores.

\subsection{CNNs models}
Our main task is to be able to distinguish whether a spectrogram has an injected signal or not, hence it's a classification problem. In this work, we made use of the most popular networks in computer vision with the state-of-the-art performance in image classification and localization, to look for the most optimal model with the highest performance regarding the problem in hand. These models are namely VGG, ResNet and DenseNet networks.

\subsubsection{VGG network}
First introduced in 2014 by \citep{https://doi.org/10.48550/arxiv.1409.1556} in the ImageNet challenge, where their submission won the first and the second places in the localization and classification tracks respectively. They built 2 models with 16 and 19 convolutional layers with 3 by 3 kernel size. We used both architectures, VGG16 and VGG19, in this work.

\subsubsection{ResNet}
Presented by \cite{https://doi.org/10.48550/arxiv.1512.03385} on the ILSVRC 2015 classification challenge, where ResNet architecture won first place. ResNet uses batch normalization, which adjusts the input layer, and identity connection between the layers to protect the network from the well-known vanishing gradient problem. ResNet has different architectures which follow the same concept but with different number of layers, for example ResNet-18, ResNet-34, ResNet-50, ResNet-101, ResNet-110, ResNet-152 etc, we made use of ResNet-101 (ResNet hereafter). 

\subsubsection{DenseNet}

In DenseNet \citep{https://doi.org/10.48550/arxiv.1608.06993}, the input does not only come from only the previous layer, but from all the preceding layers, which means all the preceding layers share their feature-maps to all subsequent layers. All layers are concatenated in DenseNet, which gives each layer an access to a collective knowledge from all the previous layers. DenseNet achieved higher accuracy compared to ResNet on ImageNet dataset. In this work, we used DenseNet-121 (DenseNet hereafter), which has 120 Convolutional and 4 AvgPool layers.

\subsection{Training}
The performance of the ML models depend largely on the quality of the training data not on the algorithms \citep{doi:10.1190/INT-2015-0044.1}, which means different algorithms should provide generally similar trend if trained on a robust and large enough data. Hence we
devoted a considerable effort to make sure our synthetic data is well representative in terms of $D_{L}$, mass distribution and balanced SNR ranges of all our BBH sources.  
We trained VGG16, VGG19, ResNet and DenseNet models with a batch size of 256, learning rate of 0.0001, 200 epochs using RSM optimizer. These parameters were chosen through running different experiments. The input layer size for all models were adjusted to $42 \times 365 \times 1$ shape size which describe the width, height and number of bands of the input image respectively (1 band denotes gray scale images). As a tradition in training ML models, our data were split into training, validation and testing sets, Table \ref{tab:no_sample} shows the total number of spectrograms used in each.

\begin{table}
\centering
\begin{tabular}{ccccc}
\hline
  Type & Number of Sample & Train & Test& Val\\
  \hline
  Injected & 25,000 & 17,000& 4000 & 4000  \\ 
  Only noise & 25,000 & 17,000 & 4000 & 4000 \\
  \hline
  Total & 50,000 & 34,000& 8000 & 8000 \\
  \hline
\end{tabular}
\caption{Total number of: injected and only noise spectrograms, training, testing and validation spectrograms.}
\label{tab:no_sample}
\end{table}

\subsection{Networks Evaluation}
\begin{figure}
\includegraphics[width=\columnwidth]{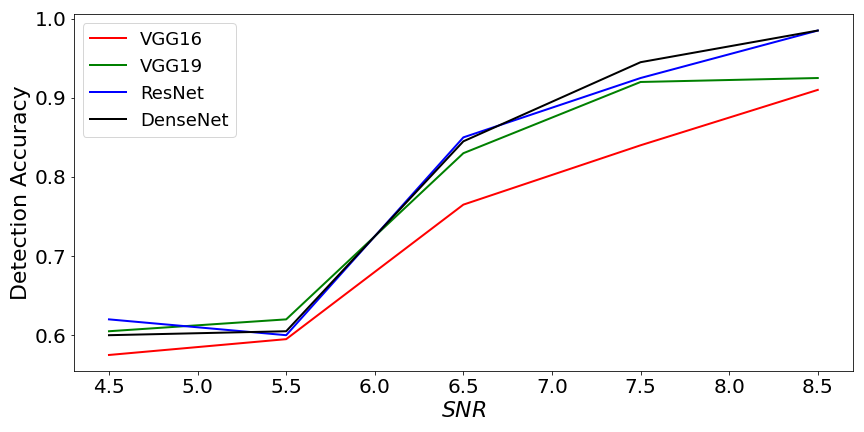}
\caption{SNR versus detection's accuracy.}
\label{fig:acc}
\end{figure}

To evaluate our trained models, a classification report was produced for each model for each SNR range where the precision, recall and F1-scores \citep{Fawcett2006} were calculated as shown in Table \ref{tab:metrics}. In terms of accuracy, DenseNet, ResNet and VGG19 showed significantly better performance over VGG16, on sources with higher SNR ranges (6-7, 7-8 and >8). DenseNet was outperformed by ResNet in SNR range 4-5 with 0.02, and VGG19 with 0.005 better accuracy score respectively. And in SNR range 6-7 by ResNet, with 0.01. In SNR range 7-8, DenseNet has better accuracy score with 0.02  over ResNet, and both has identical performance in SNR >8 of 0.985.
Same conclusion will be drawn if we compare the values of F1-score of both, ResNet and DensNet. Based on this, ResNet model was used for producing confusion matrix to further evaluate its performance. Better visualization of the accuracy for all models is shown in Figure \ref{fig:acc} where the detection accuracy is compared to all SNR ranges for each models. The general trend is exactly as expected, where higher SNR sources have higher detection accuracy. It can clearly be seen that all models can detect completely hidden signals, with different accuracy, from sources with very low SNR down to 4. 

\begin{table}
\centering
\resizebox{\columnwidth}{!}{\begin{tabular}{|c|c|c|c|c|c|c|}
Model & SNR range & class & accuracy & precision & recall & f1-score  \\\hline

\multirow{8}{4em}{VGG16} & 4-5 & injected & 0.575 & 0.913 & 0.212 & 0.344 \\ 
     &   &  noise &   & 0.554 & 0.980 & 0.708\\ \cline{2-7} & 
    
      5-6 & injected & 0.595 & 0.919 & 0.343 & 0.500 \\ 
    &  & noise &  & 0.596 & 0.970 & 0.738\\ \cline{2-7} & 
    
    6-7 & injected & 0.765 & 0.988 & 0.808 & 0.889 \\ 
    &   & noise &  & 0.838 & 0.990 & 0.907\\ \cline{2-7} & 
    
    7-8 & injected & 0.840 & 0.979 & 0.929 & 0.953 \\ 
    &   & noise &  & 0.933 & 0.980 & 0.956\\ \cline{2-7} & 
    
     > 8 & injected & 0.910 & 0.990 & 0.990 & 0.990 \\ 
    &   & noise &  & 0.990 & 0.990 & 0.990\\
    
    \hline
 
\multirow{4}{4em}{VGG19} & 4-5 & injected & 0.605 & 0.913 & 0.212 & 0.344 \\ 
     &  &  noise &  & 0.554 & 0.980 & 0.708\\ \cline{2-7} & 
    
      5-6 & injected & 0.620 & 0.900 & 0.273 & 0.419 \\ 
    &   & noise &  & 0.571 & 0.970 & 0.719\\ \cline{2-7} & 
    
    6-7 & injected & 0.830 & 0.941 & 0.808 & 0.870 \\ 
    &   & noise &  & 0.832 & 0.949 & 0.887\\ \cline{2-7} & 
    
    7-8 & injected & 0.920 & 0.967 & 0.899 & 0.932 \\ 
    &   & noise &  & 0.906 & 0.970 & 0.937\\ \cline{2-7} & 
    
     > 8 & injected & 0.925 & 0.980 & 0.970 & 0.975 \\ 
    &   & noise &  & 0.970 & 0.980 & 0.975\\
    
    \hline

\multirow{4}{4em}{RESNET} & 4-5 & injected & 0.620 & 0.604 & 0.586 & 0.595 \\ 
     &  &  noise &  & 0.598 & 0.616 & 0.607\\ \cline{2-7} & 
    
      5-6 & injected & 0.600 & 0.667 & 0.566 & 0.612 \\ 
    &  & noise &  & 0.623 & 0.717 & 0.667\\ \cline{2-7} & 
    
    6-7 & injected & 0.850 & 0.724 & 0.899 & 0.802 \\ 
    &  & noise &  & 0.867 & 0.657 & 0.747\\ \cline{2-7} & 
    
    7-8 & injected & 0.925 & 0.727 & 0.939 & 0.819 \\ 
    &  & noise &  & 0.914 & 0.646 & 0.757\\ \cline{2-7} & 
    
     > 8 & injected & 0.985 & 0.711 & 0.919 & 0.802 \\ 
    &  & noise &  & 0.886 & 0.626 & 0.734\\
    
    \hline
    
\multirow{4}{4em}{DENSNET} & 4-5 & injected & 0.600 & 0.919 & 0.343 & 0.500 \\ 
     &  &  noise &  & 0.596 & 0.970 & 0.738\\ \cline{2-7} & 
    
      5-6 & injected & 0.605 & 0.865 & 0.455 & 0.596 \\ 
    &  & noise &  & 0.630 & 0.929 & 0.751\\ \cline{2-7} & 
    
    6-7 & injected & 0.845 & 0.958 & 0.919 & 0.938 \\ 
    &  & noise &  & 0.922 & 0.960 & 0.941\\ \cline{2-7} & 
    
    7-8 & injected & 0.945 & 0.990 & 0.990 & 0.990 \\ 
    &   & noise &  & 0.990 & 0.990 & 0.990\\ \cline{2-7} & 
    
     > 8 & injected & 0.985 & 0.980 & 1.000 & 0.990 \\ 
    &   & noise &  & 1.000 & 0.980 & 0.990\\
    
    \hline

\end{tabular}}
\caption{\label{tab:metrics} Evaluation metrics for all models.}
\end{table}

\begin{figure}
\includegraphics[width=\columnwidth]{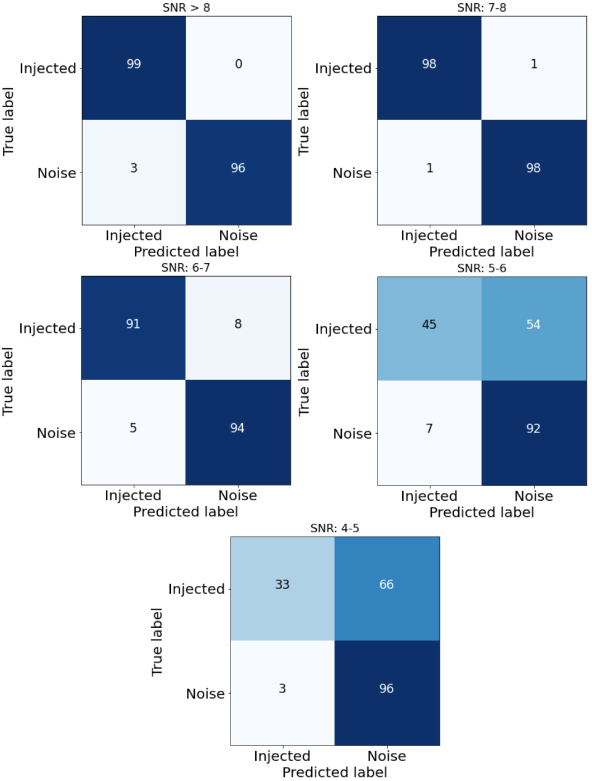}
\caption{Confusion matrix for all SNR ranges.}
\label{fig:cm1}
\end{figure}

 Confusion matrix \citep{1997RSEnv..62...77S} illustrates how accurate the model is classifying each class, injected and only noise for each SNR range in our case, through visualizing the true positive, false positive, true negative and false negative values for each class. For producing confusion matrices, we have generated 200 spectrograms (of injected signals and only noise) for each SNR range. Figure \ref{fig:cm1} displays the confusion matrices for each SNR range. For SNR >8, no only-noise spectrogram was classified as an injected one, while only one only-noise in SNR range 7-8 was miss-classified as an injection. 8 injected spectrograms were mis-classified as only-noise and 5 only-noise got classified as injected in SNR range 6-7. For SNR range 5-6, 54 injected spectrograms were classified as only-noise, while only 7 only-noise spectrograms were recognised as injected. The least performance was for SNR 4-5 sources, where 66 injected spectrograms were classified as only-noise, but only three injected spectrograms were classified as only-noise. This indicates a very good performance and means that type II error is unlikely to occur, therefore the model won't easily recognize noise as an injection.  

\section{Evaluation on ET's synthetic data}
\label{sec:eval}
It's important to evaluate our best performing model on a relatively long time series data, to check for the efficiency and practicality of a near-real time detection. To do so, we generated a mock data of five hours and 75 seconds of time series data with a signal injected randomly in every five seconds, and one hour and 15 seconds data of only noise. False Negative Rate (FNR) (also known as Type II Error rate) and False Alarm Rate (FAR) (also called false positive ratio or Type I Error Rate) are an effective metrics for measuring the performance of machine learning models applied to classification problems \citep{doi:10.1056/NEJM198810133191501}. FNR is the ratio of the total number of false negative (FN) (here the total number of undetected sources that's actually been injected) by the total number of FN and true positive (TP) (here the correct classification of injected sources that's actually injected). FNR shows how often an injected source would be miss-classified as pure noise, and mathematically is calculated as FN/FN+TP. FAR is the ratio of the total number of false positive (FP) (here the incorrect identification of only noise, i.e. classifying only noise as injected) by the total number of FP and true negative (TN) (here the total number only noise). FAR shows how often the model incorrectly classifies an only noise spectrogram as an injected, and FAR mathematically is calculated as: FP/(FP + TN).

\begin{figure}
\includegraphics[width=\columnwidth]{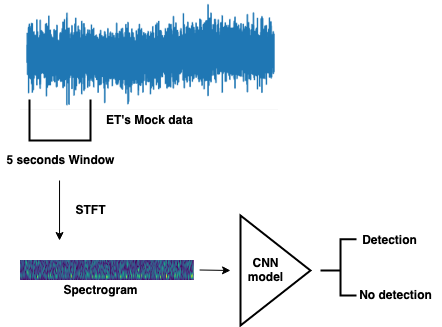}
\caption{Inferencing on ET's synthetic data}
\label{fig:pipe}
\end{figure}

Each segment (hour and 15 seconds) in the injected mock data contains 723 signals, which belong to specific SNR range. For the detection and calculation of FNR on this data, we built a pipeline where an sliding window of 5 seconds size will be moved through the data to capture a 5 seconds chunk each time. Then a spectrogram will be produced for each chunk and fed into the model for inference. Figure \ref{fig:pipe} displays the flow chart of the ET's mock data inference process. 
Table \ref{tab:longts} shows the report for each hour, where the number of detected sources is compared to the total number of injected signals, and FNR is calculated for each hour. 722 sources with SNR >8 out of 723 total sources were detected, with only 0.001 FNR. For sources with 7-8 SNR range, 713 sources were detected with an FNR of 0.014, while sources with 6-7 SNR range has an FNR of 0.05 with 687 sources detected. Sources with 5-6 and 4-5 SNR range have lower FNR as expected, with 336 and 208 sources were detected respectively. FAR was calculated on the only noise 1 hour and 15 seconds data, where 24 nonexistent sources were detected with  0.033 FAR. This confusion is due to the training on signals with lower SNR. For instance, sources with SNR of 4 to 5 -on the spectrogram- would look identical to a pure noise once injected. When training an CNNs model on sources with SNR >= 8, the model won't confuse a pure noise with signals and therefore less FAR. The confusion was very limited with only 0.033 FAR.

\begin{table}
\centering
\begin{tabular}{|c|c|c|c|c|}
Hours & SNR range & Total & Detected & FNR/FAR  \\\hline
1st & >8 & 723 & 722 & 0.001  \\
2nd & 7-8  & 723 & 713 & 0.014 \\
3rd & 6-7 & 723 & 687 & 0.050  \\
4th &  5-6 & 723 & 336 & 0.535  \\
5th & 4-5 & 723 & 208 &  0.712 \\\hline
6th & Noise only & 0 & 24 & 0.033  \\ \hline
\end{tabular}
\caption{\label{tab:longts} Detection evaluation on ET's synthetic data.}
\end{table}

The scan of each hour from the data took 4 minutes and 57 seconds on an Core i7 MacBook Pro (with 16 GB memory, 2667 MHz DDR4 and 2.6 GHz processor) without parallelizing the code. This indicates that, the use of CNNs for merger detection is computationally efficient and can be significantly improved.  

To further explore and evaluate the performance of the model, we plot the SNR, $D_{L}$ and chirp mass $M$ of the detected and undetected BBH sources. The chirp mass governs the motion and the waveform of the binary during the in-spiral phase \citep{Bulik_2003}, and it depends on the individual masses $m_1$ and $m_2$ of the binary component, calculated as $M = (m_1 + m_2)^{-1/_5} (m_1 \times m_2)^{3/_5}$ \citep{PhysRev.131.435}.

\begin{figure}
\includegraphics[width=\columnwidth]{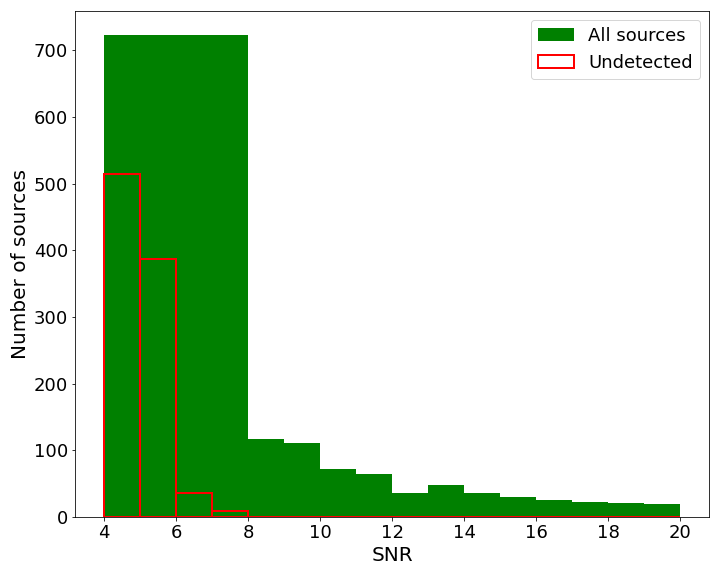}\\
\includegraphics[width=\columnwidth]{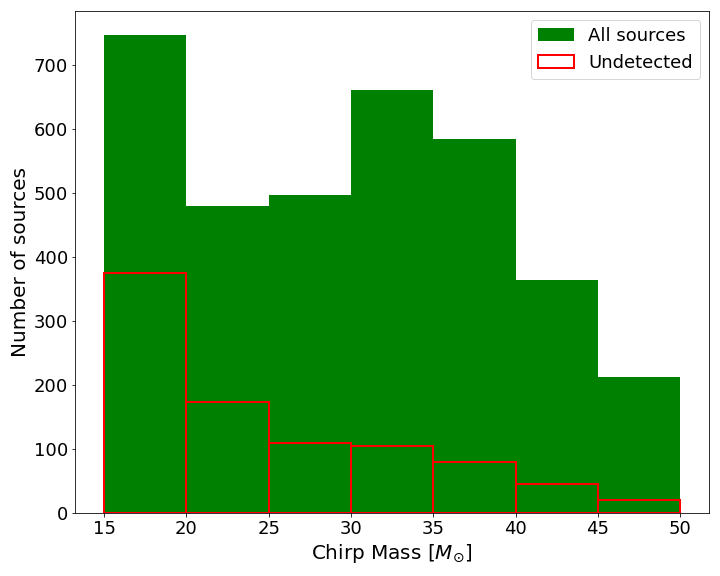}
\caption{Top: SNR of all BBHs sources and the undetected ones. Bottom: Chirp mass of all BBHs  sources and the undetected ones.}
\label{fig:detection_snr_m}
\end{figure}

\begin{figure}
\includegraphics[width=\columnwidth]{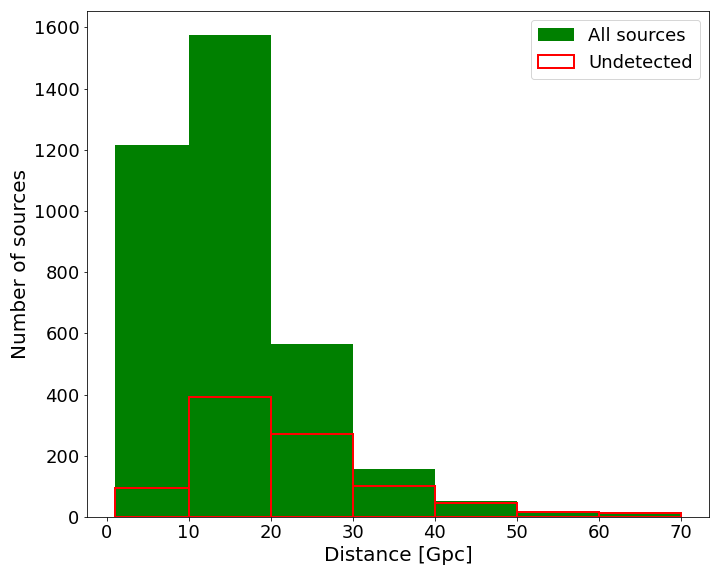}
\caption{$D_{L}$ of the all BBHs sources and the undetected ones.}
\label{fig:detection_dist}
\end{figure}

Figure \ref{fig:detection_snr_m} and Figure \ref{fig:detection_dist} displays the SNR, $M$ and $D_{L}$ associated with the detected and undetected BBH sources respectively. Detected sources range in $D_{L}$ from $125$ - $60$ Gpc, SNR from 4 - 168, and chirp mass from  15 - 56 \(M_\odot\). Undetected sources have a left skewed distribution for chirp mass and maximum $D_{L}$ of 117,5 Mpc with SNR range between 4 to 8.8. Most undetected sources have lower SNR and mass with larger $D_{L}$. Minimum SNR detected source has an SNR of 4 with $D_{L}$ of 17,9 Mpc and 23.6 $M$, while the maximum $D_{L}$ of a detected source has SNR of 4.3 with a $D_{L}$ of $60$ Gpc and $M$ of 35 \(M_\odot\).

\section{Conclusions and future work}
\label{sec:con}
VGG16, VGG19, ResNet and DenseNet were trained and tested for the detection of BBHs merger from ET's synthetic data with five SNR ranges. Without performing data whitening or applying bandpass filtering, ResNet has significantly better performance with an average accuracy of 85\%.
On testing data. ResNet model was able to detect BBHs sources with SNR of 8 or higher with 98.5\% accuracy, and with 92.5\%, 85\%, 60\% and 62\% accuracy for sources with SNR range of 7-8, 6-7, 5-6 and 4-5 respectively. For 4-5 SNR range. ResNet model was able to detect a BBH's merger at 60Gpc with 4.3 SNR and a chirp mass of 35 \(M_\odot\). It was also shown that, the use of CNN for merger detection is computationally efficient, in our case, scanning a one hour of data only took 4 minutes and 57 seconds. This can be significantly improved, which makes our method -compared to matched filtering- more efficient for near-real-time detection. Future work will focus on spinning BBHs and BNSs sources, injecting glitches and consider the overlapped sources scenario.

\section*{Acknowledgements}
The authors thank Prof. Piotr Gawron and Dr. Josiah Ensing for their helpful discussions and reviews. This work has been supported by the European Union’s Horizon 2020 research and innovation programme under grant agreement No 952480 (DarkWave project), and from the International Research Agenda Programme AstroCeNT \mbox{(MAB/2018/7)} funded by the Foundation for Polish Science (FNP) from the European Regional Development Fund. Additionally it made use of the computational cluster available at Nicolaus Copernicus Astronomical Center.

\section*{Data Availability}
Data used in this work -including time-series and spectrograms- can be shared on a reasonable request, please contact the author to request the data.



\bibliographystyle{mnras}
\bibliography{bbh_detection} 








\bsp	
\label{lastpage}
\end{document}